\documentclass[aps,prb,twocolumn,reprint,superscriptaddress,english]{revtex4-1}
\usepackage[T1]{fontenc}
\usepackage[latin9]{inputenc}
\usepackage{graphicx}
\usepackage{float}
\usepackage{wrapfig}
\usepackage[dvipsnames]{xcolor}
\usepackage[linktocpage=true,colorlinks=true,pdfborder={0 0 0},linkcolor=blue,
citecolor=blue,filecolor=yellow,urlcolor=blue,bookmarks,pdfauthor={},]{hyperref}
\usepackage{amsmath}
\usepackage{gensymb}
\usepackage{lastpage}
\usepackage{enumerate}
\usepackage{mathrsfs}
\usepackage{xcolor}
\usepackage{graphicx}
\usepackage{listings}
\usepackage{hyperref}
\usepackage{mathtools}
\usepackage{physics}
\usepackage{amsmath}
\usepackage{dblfloatfix}

\newcommand{\NEXCL}{Nevada Extreme Conditions Laboratory, University of Nevada, Las Vegas, Las Vegas, Nevada 89154, USA}
\newcommand{\LVPhys}{Department of Physics \&{} Astronomy, University of Nevada Las Vegas, Las Vegas, Nevada 89154, USA}

\begin{document}
\author{Andrew~L.~Cornelius}
\affiliation{\LVPhys}
\email{andrew.cornelius@unlv.edu}
\author{Keith~V.~Lawler}
\affiliation{\NEXCL}
\author{Ashkan~Salamat}
\affiliation{\LVPhys}
\affiliation{\NEXCL}

\date{\today{}}

\title{Understanding Hydrogen Rich Superconductors: Importance of Effective Mass and Dirty Limit}

\begin{abstract} 
 A class of hydrogen-rich (H-rich) systems, consistent with Type II superconductivity, are known to have very high superconducting transition temperatures $T_{c}$ and upper critical magnetic field $B_{c2}$ values (up to 288 K and 222 T respectively).  
 By looking at all the experimental H-rich superconductors reported to date most are in the dirty limit, with only the highest $T_{c}$ values being in the crossover to the clean regime.
 In this framework, there is a clear understanding of some previous unexplained behaviors:  (1) A maximum in $B_{c2}$ as a function of $T_{c}$; (2) a clear change in slope in $T_{c}$ as a function of pressure in CSH; (3) in zero magnetic field the width of the superconducting transition decreasing with increasing $T_{c}$; and (4) in applied magnetic field the slope of the superconducting width versus field decreasing with increasing $T_c$. Ginzburg-Landau-Abrikosov-Gorkov (GLAG) theory is used to explain all four of these effects within a framework where increasing $T_{c}$ is related to increases in both the electron effective mass and scattering length.

\end{abstract}  
\maketitle 

\section{Introduction}

The route to ambient conditions superconductivity will most likely be achieved from a class of H-rich systems that behave as hydrogen dominant alloys, mimicking metallic hydrogen but at significantly more modest densities.\cite{RN39,RN40} The first report of room temperature superconductivity by \citet{RN1} in carbonaceous sulfur hydride (CSH) was measured at 288\,K and 268\,GPa. Their electrical transport measurements revealed a very sharp transition from the normal to zero-resistance state as a function of both $T_{c}$ and magnetic field, especially in relation to previous reports on other Type-II superconductors YBCO\cite{RN14} and MgB$_2$ (see Table II). The relative superconducting transition width, $\Delta T_{c}/T_c$, of CSH decreases in both zero and applied magnetic fields with increasing $T_{c}$. Understanding the limits in which these H-rich superconductors lie is key to understanding their properties including the transition widths.

There are many ways that the original Ginzburg-Landau equations\cite{RN4} have been expanded by the ideas of Abrikosov-Gorkov (GLAG)\cite{RN5} to include microscopic theory, discussed here using the notation of Schmidt.\cite{RN6} In cgs units, the difference in free energy between the superconducting and normal state is G$_s$--G$_n$=G$_{pair}$+G$_K$+G$_H$, where G$_{pair}$ is the condensation energy of the Cooper pairs, G$_K$ is the kinetic energy of the electrons and G$_H$ is the energy due to magnetic field expulsion. In terms of the superconducting wave function $\Psi$ which is the order parameter:
\begin{equation} \label{eq:1}
G_{pair} = \alpha \vert t \vert\vert\Psi\vert ^2 + 
\frac{\beta}{2}\vert\Psi\vert^4.
\end{equation} 
Here $t=1-T/T_{c}$ is the reduced temperature that goes from 1 $\rightarrow$ 0 for temperatures below the critical superconducting transition temperature, $T_{c}$. 

An important parameter is the coherence length, $\xi$, which is the extent of the wavefunction $\Psi$. It is also related to the s-wave BCS superconducting Cooper pair size, and it is the range of the superconducting vortex core in the mixed region in a magnetic field. For consistency, the notation for the given GLAG parameters are their zero temperature values, and their temperature dependence will be given by multiplying these terms by the appropriate quantity as done in Equation~\ref{eq:1}. Typically, applications of this theory are given in either the clean or dirty limit. In the clean limit $\xi=\xi _{GL}$. However, in the dirty limit where the electron mean free scattering length $\ell$ is much smaller than $\xi_{GL}$, one needs to use $\xi=(\ell\xi_{GL})^{1/2}$. There is very little discussion in the literature about the use of these equations in the crossover region between clean and dirty. To bridge the regions, one should use the equation $\xi^{2}=\xi_{AV}\xi_{GL}$ where $1/\xi_{AV}=1/\ell+1/\xi_{GL}$. Another important parameter is the upper critical field where a magnetic field is no longer expelled from the superconductor, $B_{c2}=\Phi_{0}/2\pi\xi^2$ where $\Phi_{0}$ is the flux quantum $hc/2e$. As $B_{c2}$ is a measurable quantity, it is used to experimentally determine $\xi$, though it is often incorrectly reported in the literature as $\xi_{GL}$ implying the clean limit. In typical experimental units, the relationship between $\xi$ and $B_{c2}$ is
\begin{equation}\label{eq:2}
\xi\left(nm\right)=\frac{17.9}{\sqrt{B_{c2}\left(T\right)}}.
\end{equation}

Moving from the dirty to the clean limit has a profound impact on the superconducting parameters of a system. A recent survey of results within the GLAG model on A15 superconducting Nb$_{1-\beta}$Sn$_{\beta}$ (0.18<$\beta$<0.26) materials display this impact.\cite{RN3} At $\beta$=0.18 Nb$_{1-\beta}$Sn$_{\beta}$ is clearly in the dirty limit ($\ell\gg\xi_{GL}$). As $\beta$ is increased, linear behavior is observed in the electron effective mass $m^*$ both from fits to models to obtain experimental $T_{c}$ values and electronic Sommerfeld coefficient from heat capacity measurements.\cite{RN13} Figure 1 shows how the various GLAG parameters vary as one increases $m^*$ by increasing $\tau$. It is important to note that $m^*\propto\beta\propto T_{c}$, so $\tau$ can represent either $m^*/m_0^*$, $\beta/\beta_{0}$ or $T_{c}$/$T_{c0}$
as the dependent dimensionless variable with an appropriate choice of a reference point. 

\begin{figure}[h]
    \includegraphics[width=\columnwidth]{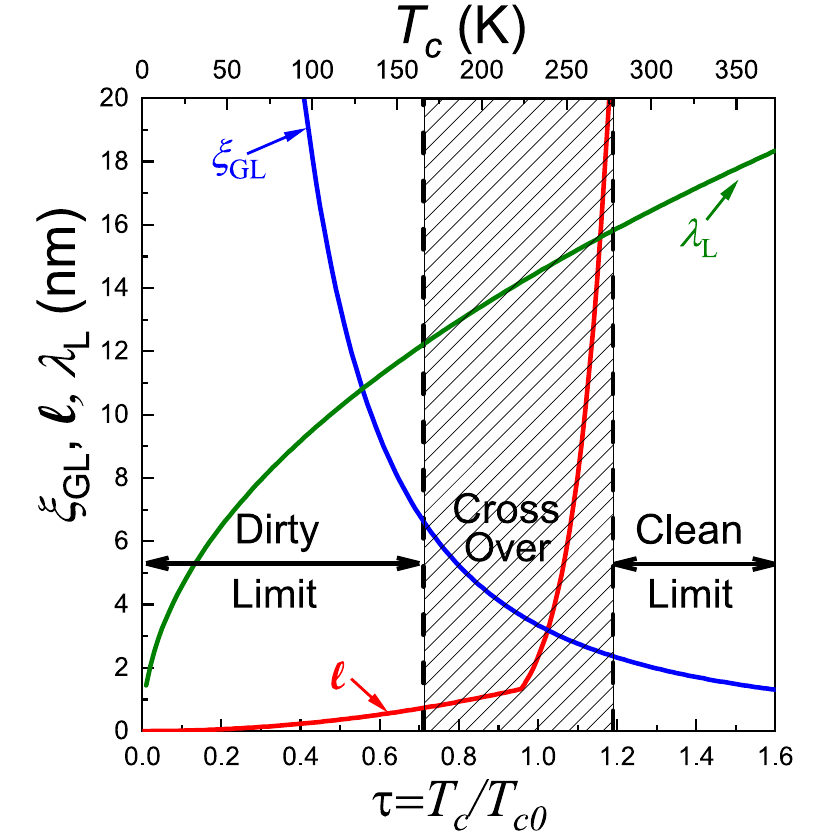}
    \caption{Various physical parameters from the Ginzburg-Landau-Abrikosov-Gorkov (GLAG) theory applied to H-rich superconductors as a function of superconducting transition temperature $T_{c}$ (raw value on top x-axis and normalized on the bottom x-axis). $\xi_{GL}$ is the coherence length in the GLAG model in the clean limit. $\ell$ is the mean electron scattering length. $\lambda_{L}$ is the Landau magnetic field penetration depth. }
    \label{fig:1}
\end{figure}

\begin{figure}[h]
    \includegraphics[width=\columnwidth]{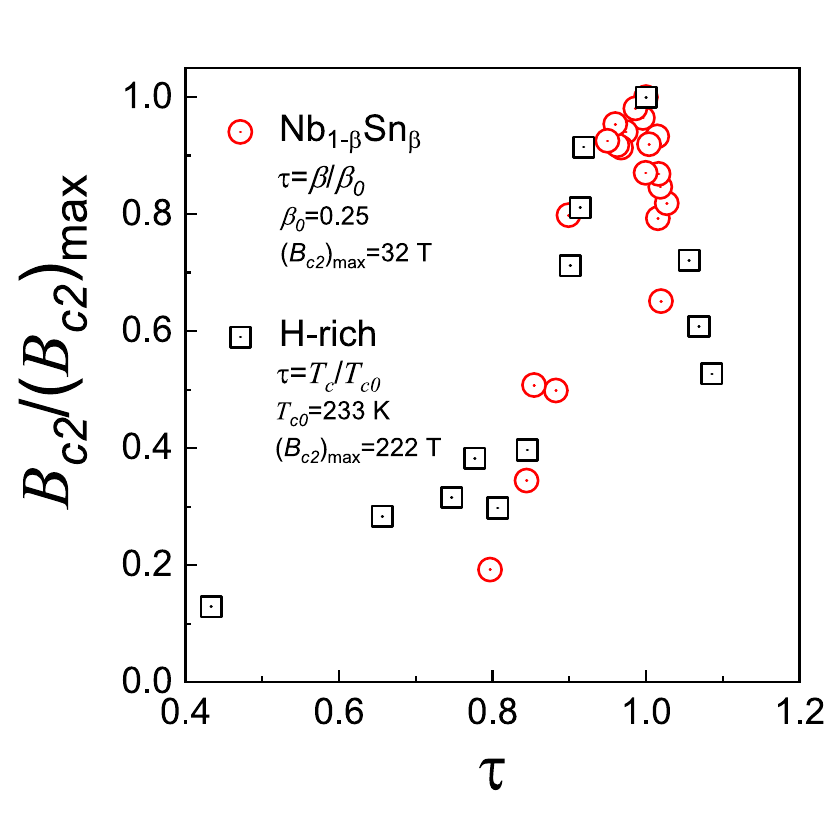}
    \caption{
The upper critical magnetic field $B_{c2}$ versus normalized control parameter $\tau$ for A15 Nb$_{1-\beta}$Sn$_{\beta}$ and H-rich materials. The upper critical magnetic field $B_{c2}$ is normalized by its maximum value chosen to be at $\tau=1$. $\tau=\beta/\beta_{0}$ for  Nb$_{1-\beta}$Sn$_{\beta}$ and $T_{c}$/$T_{c0}$ for H-rich superconductors.}
    \label{fig:2}
\end{figure}

Over the entire range $\xi_{GL}$ and the London penetration depth, $\lambda_{L}$, are assumed to follow their expected GLAG $\tau^n$ behavior with $n=-2$ and $+1/2$ respectively. The actual penetration depth can be found using $\lambda_{L}$ in the appropriate clean/dirty limit. Initially $\ell$ increases in a steady power law $\ell\sim\tau^{n}$ manner. As $\ell$ and $\xi_{GL}$ become nearly equal, a superconductor is in the crossover region between the dirty and clean limits, where one must use $\xi_{AV}$ rather than just $\ell$ (dirty) or $\xi_{GL}$ (clean). Further moving to the right in Figure~\ref{fig:1} leads to a much steeper increase in the observed $\ell$ deduced from $B_{c2}$ and electric transport measurements. The behavior depicted in Figure~\ref{fig:1} has implications for many measurable quantities. One is when $\ell\sim\xi_{GL}$, the value $B_{c2}$ can be found to have a maximum corresponding to a minimum in $\xi$ per Equation~\ref{eq:2}. Figure~\ref{fig:2} shows a plot of $B_{c2}$ normalized by its maximum value plotted versus $\tau=\beta/\beta_{0}$ to highlight this maximal value for $B_{c2}$.

For H-rich materials, it is necessary to orient them within the framework of Figure~\ref{fig:1}. For our analysis of H-rich superconductors, $B_{c2}$ is calculated using the Werthamer-Helfand-Hohenberg (WHH) formalism for consistency. Thus, the published values employed in this analysis are used directly if they were calculated in the WHH formalism, if not they are estimated using the published data and the WHH equation, $B_{c2}$\,=\,--0.69\,$T_{c}$\,$B_{c2}\prime$ where $B_{c2}\prime$ is the temperature derivative of $B_{c2}$ at $T_{c}$. One difficulty fitting H-rich superconductors to the GLAG model is that all of the measurements are done at very high pressures in a diamond anvil cell (DAC) limiting the number of experimental probes. While the A15 compounds have the advantage of access to the full suite of thermodynamic measurements at ambient pressure, most H-rich measurements in the DAC are electrical transport. While $B_{c2}$ and sometimes $\ell$ can be extracted from these measurements, other quantities are elusive.

A recent work determining the GLAG parameters for two H-rich superconductors, H$_3$S and LaH$_{10}$ with $T_{c}$ values near $T_{c0}$=233\,K were able to perform quantifiable magnetic measurements in a DAC leading to a measurement of the Ginzburg-Landau parameter $\kappa$ that is used to divide Type-I and Type-II superconductors with $\kappa^2 >1/2$ leading to Type-II superconductivity. $\kappa$ is the ratio of the London penetration depth $\lambda_{L}$ to $\xi_{AV}$. Table I shows a summary of the results. These results are for two Type-II superconductors with large $B_{c2}$ values near the $B_{c2}$ maximum and lie in the crossover region where $\ell\sim\xi_{GL}$, which is exactly the case for both materials. The London penetration depth $\lambda_{L}$ was taken as the midpoint of the range of values estimated for LaH$_{10}$ in Table I. The clean limit is found where $\ell\gg\xi_{GL}$ and $\ell\sim\lambda_{L}$.

In the H-rich superconductors, like the A15 materials, $m^*\sim\,T_{c}$. Figure~\ref{fig:2} shows $B_{c2}$ for the H-rich superconductors normalized by its maximum value plotted versus $T_{c}$/$T_{c0}$ for all of the high field data on different H-rich superconductors (where the 0 reference point is chosen as the value where $B_{c2}$ reaches its maximum). The data are in Table III. The overlap is striking and gives strong evidence that Figure~\ref{fig:1} should indeed be applicable to the H-rich superconductors. This implies that many of the H-rich superconductors will lie in the dirty region $(\ell\ll\xi_{GL})$ and that the maximum of $B_{c2}$ occurs in a cross over region between the dirty and clean limits.

For CSH, $B_{c2}$ decreases as $T_c$ increases in contrast to lower $T_c$ H-rich systems leading to a maximum in $B_{c2}$ as a function of $T_{c}$. This maximum can be explained by a crossover from the dirty to clean regime akin to what is represented in Figure~\ref{fig:1}. The scenario with CSH is very similar to that observed in A15 Nb$_{1-\beta}$Sn$_{\beta}$ where a change in stoichiometry, rather than pressure, leads to many of the same results.\cite{RN3} Herein, we will explain that the decreasing superconducting transition width of CSH is what is expected over the entire the dirty regime, and set forth an explanation for the transition width in H-rich conductors in terms of GLAG theory. Likewise, we will demonstrate that moving into the crossover regime explains the change in the slope of the pressure dependent measured $T_{c}$ values for CSH. Lastly, we will rationalize the changes in field dependent transition width in CSH and other H-rich superconductors in this context.

\subsection{Moving Towards the Dirty Limit in the H-rich Materials}

To continue the analysis,  $\ell$ and $\xi_{GL}$ will be plotted on log--log plots to look for general $\tau^n$ dependence over wide ranges representative of Figure 1. This is obviously going to be an overly simplistic view. For example while the actual value of $T_{c}$ versus $m^*=1+\lambda_{e-p}$ for different theories are not linear, it is close enough for general scaling arguments ($\lambda_{e-p}$ is the electron-phonon coupling).\cite{RN7} These parameters are: $T_{c}$/$T_{c0}=\tau$, $\xi_{GL}/(\xi_{GL})_{0}=\tau^{-2}$, $\lambda_{\ell}/\lambda_{L0}=\tau^{1/2}$ and $\rho_{n}/\rho_{n0}=(\ell/\ell_{0})^{-1}=\tau^{n}$.\cite{RN8} Note that there is no $m^*$ dependence on $\ell$ which needs to be determined by experiment.

H$_3$S is the natural continuation point as $R(T)$ data exists for a large region $T_{c}$<$T_{c0}$ where one moves entirely to the dirty limit. \cite{RN9}  Finding the dependence of $\ell/\ell_0$ is then critical for determining both the dependence of the GLAG parameters comparing to Figure 1 and how the superconducting transition width will depend on $\tau$. One of the first things that stands out is the normal state resistance decreases dramatically with increasing pressure while $T_{c}$ increases (see inset of Figure 3). This implies that $\ell$ increases with $T_{c}$.  To obtain $\ell$ one must know the conduction electron density $n$ and the sample thickness $t$. With the high pressure values being approximately $n=8.4\times10^{22}\,cm^{-3}$ and $t$=2\,$\mu m$ respectively,\cite{RN9, RN10} $\ell$ can be estimated from resistance measurements by the equation:

\begin{equation}\label{eq.3}
    \ell\left(nm\right)=\frac{1.27\times10^{4}\left(n^{-\frac{2}{3}}\right)t}{\rho_{\Omega-cm}}=\frac{0.235}{R_\Omega}
\end{equation}

For consistency and simplicity, the value of $R$ will be taken just above $T_c$ introducing some uncertainty but typically is small related to other effects. The values of $R$ above $T_c$, $R(T_c)$, is plotted versus $T_c$ in the Figure 2 inset. Note that no pressure dependence is used for the $n$ and $t$ values. The uncertainty in the absolute value of $\ell$ is around 50\%. This will not change the final discussion (namely the power law dependence of physical properties will be identical) of results, and the relative change uncertainty will probably lie mostly in the aforementioned pressure dependent values of $n$ and $t$ which should be small at very high pressures. The values of $\ell$ determined from Equation 3 using the $R(T_c)$ data are shown in Figure 3 with $\ell$ increasing as $\tau^{1.89\pm0.13}$.

The values of $\xi_{GL}$ shown in Figure 3 were obtained from the equation $\xi=\sqrt{(\ell\xi_{GL})}$. The value of $\xi$ was estimated from $B_{c2}$ measurements on H$_3$S with a fit to the data for $\tau$<0.8 gives $\xi$=1.95$\tau^{-0.31}$. The value of $\xi_{GL}$ are found to decrease at a rate of $\tau^{-2.18\pm0.10}$. The exponent found for $\xi_{GL}$ is remarkably close to the expected -2 value giving credibility to the GLAG formalism used. The data look similar to the expected scenario laid out in Figure 1. Notably, the extrapolated values of $\ell$ and $\xi_{GL}$ become equal for $\tau$ slightly greater than 1. To estimate how close the data lie to the crossover region, the largest temperature where $\ell$ is determined here for H$_3$S is $\tau$=0.64 where $\xi=0.89 \ell$, is right at the crossover region defined here as going from $\xi_{AV}=0.9 \ell$ on the dirty side and $\xi_{AV}=0.9 \xi_{GL}$ on the clean side.

\begin{figure}[h]
    \includegraphics[width=\columnwidth]{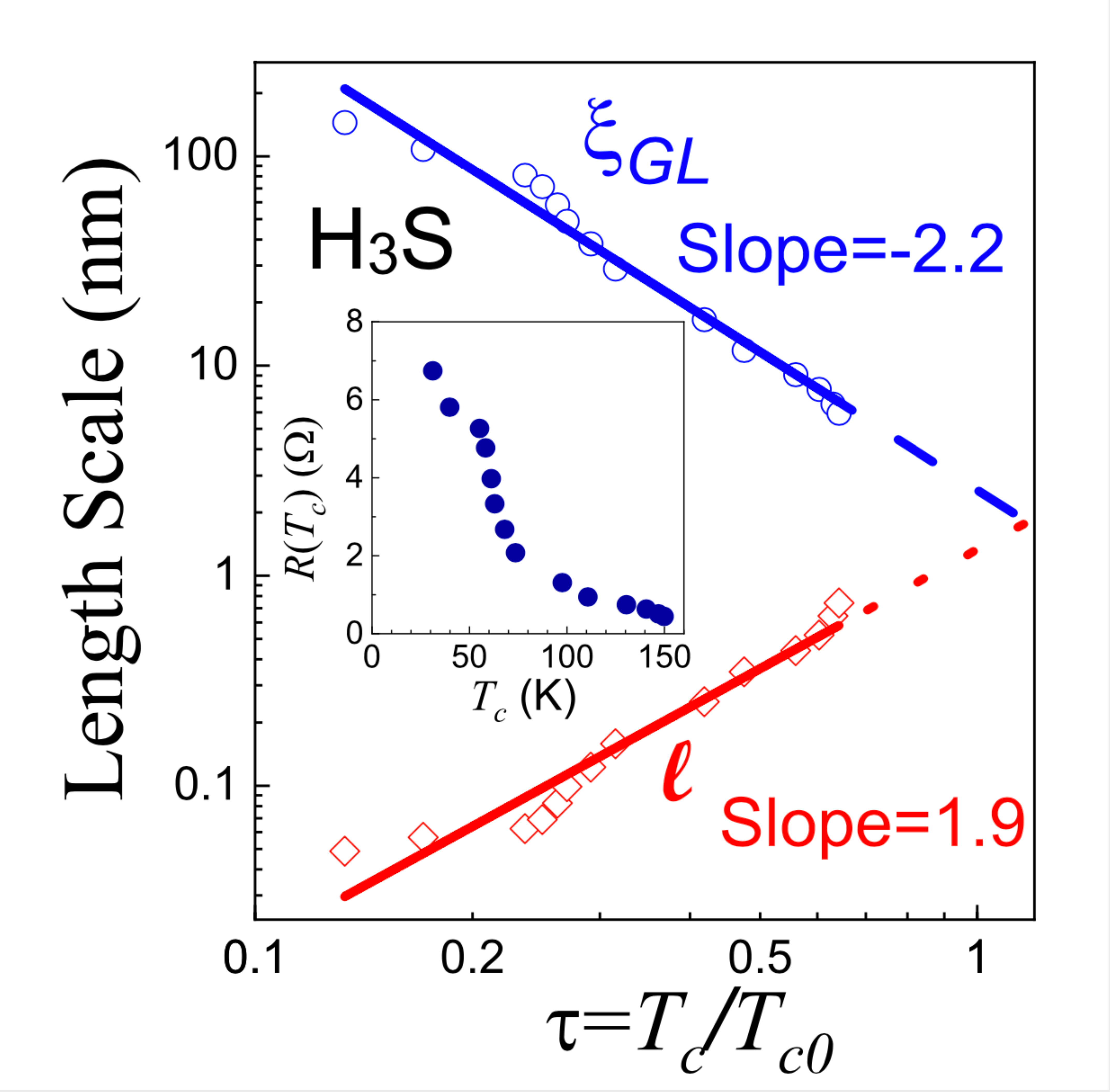}
    \caption{Calculated values of the mean free electron path $\ell$ and Ginzburg-Landau coherence length $\xi_{GL}$ from H$_3$S data. The inset shows the value of the resistance just above $T_c$, $R(T_c)$ versus $T_c$.\cite{RN9}}
    \label{fig:3}
\end{figure}

\subsection{Looking at the Crossover Region}

Moving on to $\tau>0.6$ will put a H-rich superconductor into the crossover region. For simplicity and to stay within the ideas of the GLAG model, the exponent for $\xi_{GL}$ will be fixed at -2. Fitting the data gives $\xi_{GL}=3.35\tau^{-2}$ which will be used over the entire range of $\tau$ values. Likewise, fixing the $\ell$ exponent at 2 gives the fit  given $\ell=1.45\tau^2$.

\begin{figure}[h]
    \includegraphics[width=\columnwidth]{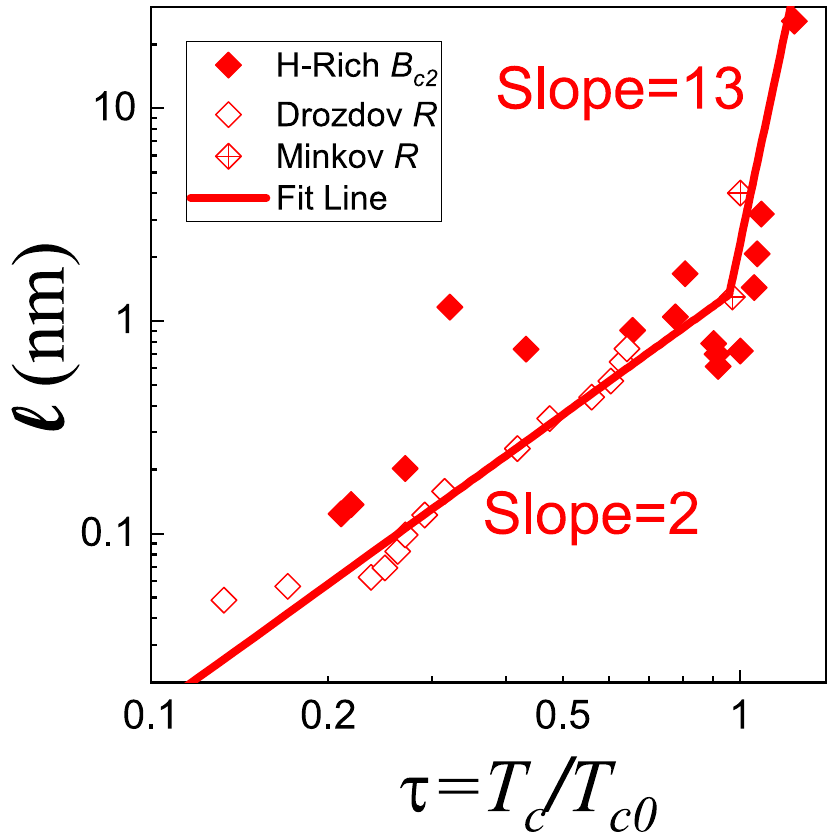}
    \caption{
Mean free electron path for H-rich superconductors as a function of $T_{c}$/$T_{c0}$. This includes resistance data from Figure 3 and Table II (2 points only). Calculated values of the mean free electron path $\ell$, using the $B_{c2}$ data (as outlined in text) for all H-rich materials. The sharp upturn is fit using all points in the plot for $T_{c}/T_{c0}\geq0.96$. }
    \label{fig:4}
\end{figure}

Another way to obtain the electron mean free path will be to look at the $B_{c2}$ data. Recall $B_{c2}$ gives the value of $\xi=(\xi_{AV}\xi_{GL})^{1/2}$ where $1/\xi_{AV}=1/\ell+1/\xi_{GL}$. Figure 4 shows the extrapolated fits for $\xi_{GL}$ and $\ell$ along with the experimental calculation of $\ell$ from resistance (Fig. 3) including the two points from transport measurements in Table I and $B_{c2}$ measurements. The values of $\ell$ from the $B_{c2}$ data qualitatively track the resistance data. The main result of Figure 4 is the sharp increase of $\ell$ that starts in the region where $B_{c2}$ reaches a maximum. Fitting all of the data in Figure 4 for $\tau>1$ gives a $\tau^{13}$ dependence The $\tau^2$ fit to $\ell$ region is extended up until its intersection with the $\tau^{13}$ line which occurs at $\ell$=1.34 nm. This gives $\ell=2.23\tau^{13}$ nm for $\tau$>0.96. At $\tau$=0.96, $\xi_{GL}$=4.64 nm which is still larger than $\ell$. However, a slight increase in $\tau$ leads to a large increase in $\ell$ and $\ell=\xi_{GL}$ at $\tau$=1.03. The numeric values displayed in Figure 1 have all been calculated and will be used for the following discussion.

Figure 5 shows the overall dependence of $\ell$ and $\xi_{GL}$ normalized by the $\xi_{AV}$ values as a function of $\tau$. The two values reach their 0.5 values, where they are equal, at $\tau=1.025$. Recall that the clean limit corresponds to $\xi_{AV}$=$\xi_{GL}$ while the dirty limit gives $\xi_{AV}$=$\ell$. While somewhat arbitrary, if $\xi_{AV}$/$\ell$=0.9 and $\xi_{AV}$/$\xi_{GL}$ =0.9 are chosen for the dirty and clean limits, the crossover region extends from 0.71<$\tau$<1.19. This gives quantitative values for the crossover region as also shown in Figure 1 (showed by the hatched region in both figures).

Looking at the GLAG theory applied to BCS,\cite{RN6} $T_{c}$ in the dirty limit, takes the form $(T_{c})_\text{dirty}= \alpha\tau^{n}\times(\xi_{GL}\ell)^{-1/3}$, where the $\tau^{n}$ dependence is due to physical terms like the effective mass and the density of states at the Fermi level and $\alpha$ is a constant. $\alpha$ is slightly different for the clean and dirty limits but will be assumed to be the same for discussion. The product of $\xi_{GL}\ell$ in the dirty limit is independent of $T_{c}$ so $(T_{c})_\text{dirty}= \alpha\tau^{n}$ meaning that $n=1$. The relationship between the dirty and clean limits can be written as $(T_{c})_{\text{clean}}/(T_{c})_{\text{dirty}}=(\ell/\xi_{GL})^{+1/3}$ which leads to the crossover approximation of 
$T_{c}$= $T_{c0}\alpha\tau(\xi_{AV}/\xi_{GL})^{+1/3}$ so $T_{c}$ will no longer be directly proportional to $\tau$ due to the $\tau$ dependence of the $\xi$ terms not cancelling each other.

\begin{figure}[t!]
    \includegraphics[width=\columnwidth]{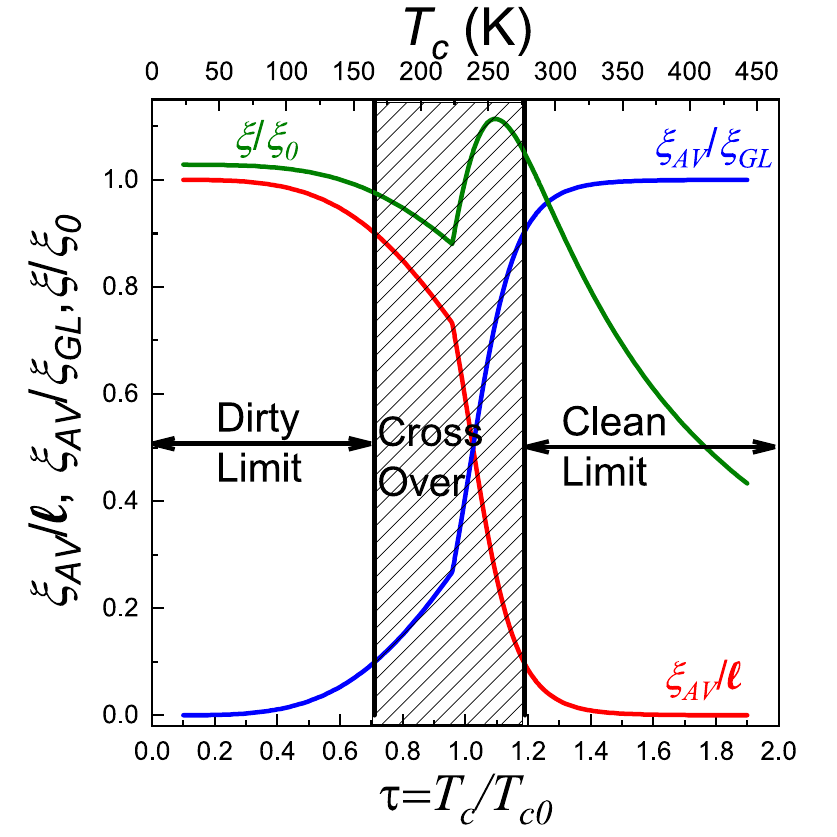}
    \caption{
Plot of $\xi_{AV}/\ell$ and $\xi_{AV}/\xi_{GL}$ as a function of $\tau$=$T_{c}$/$T_{c0}$ for the aggregate H-rich system data. $\ell, \xi_{AV}$, and $\xi_{GL}$ are the electron mean scattering length, average coherence length, and Ginzburg-Landau correlation length. The average coherence length is given by $1/\xi_{AV}=1/\ell+1/\xi_{GL}$. The dirty limit ($\ell\ll\xi_{GL}$) here is defined as $\xi_{AV}/\ell\leq0.9$. The clean limit ($\ell\gg\xi_{GL}$) here is defined as $\xi_{AV}/\xi_{GL}\geq$0.9. The cross over region is the area between the clean and dirty limits. Note that $T_{c0}$, the superconducting transition temperature corresponding to the maximum in $B_{c2}$ value clearly lies within the crossover region. }
    \label{fig:5}
\end{figure}

Turning to the CSH data of \citet{RN1}, the $T_{c}$(P) behavior shows a clear break in slope at high pressure around $T_{c}$/$T_{c0}$=0.96 where $\ell$ drastically changes slope. The data for CSH lie almost entirely in the crossover region and give a good opportunity to test the application of GLAG parameters in this region where the $T_{c}\sim m^*$ scaling will begin to break down. Looking back to the analysis of De Silva, pressure was used as the driving force to increase $m^*$ and therefore $T_{c}$. This was done by fitting $T_{c}/T_{c0}=10^{K\Delta P}=\tau_{P}$ (De Silva used $e^{K\Delta P}$ but log base 10 is used here to keep analysis consistent) where K is a constant. $\Delta P=P-P_{0}$ where $P_0$ is the pressure where $T_{c}$=$T_{c0}$=233\,K, the temperature the hydrides all appear to aggregate around as seen in Figure~\ref{fig:2}. The subscript $P$ here is for pressure to not confuse with the scaling definition that $\tau \equiv T_{c}/T_{c0}$. Note that $\tau\sim\tau_{P}$ in the dirty region, but while $\tau_{P}$ should continue to smoothly increase with pressure, the behavior of $\tau$ should be more complicated. Fitting the data from $T_{c}$/$T_{c0}$=0.69 to 0.85 which is on the dirty side of the crossover limit (and well before $\ell$ has the large increase) gives K=0.00173\,GPa$^{-1}$. A plot $T_{c}$ versus $\tau_{P}$ should have a slope of $T_{c0}C_{0}(\xi_{AV}/\xi_{GL})^{+1/3}$ where $\alpha$ is the only fitting parameter. Using the emperical values from Figure 5, the best fit occurs using $\alpha$=0.70. Figure 6 shows CSH data of Snider along with this fit over an extended range of $\tau_{P}$. The fit clearly models the data for all $\tau_{P}$ values including the change in slope of the $T_{c}$ data due to the sharp increase in $\ell$ in this region. $\tau_{P}$=1.2 is just short of the clean region, and by $\tau_{P}$=1.3 extrapolations will give the expected $\tau_{P}^{7/3}$ clean limit behavior.

\begin{figure}[b!]
    \includegraphics[width=\columnwidth]{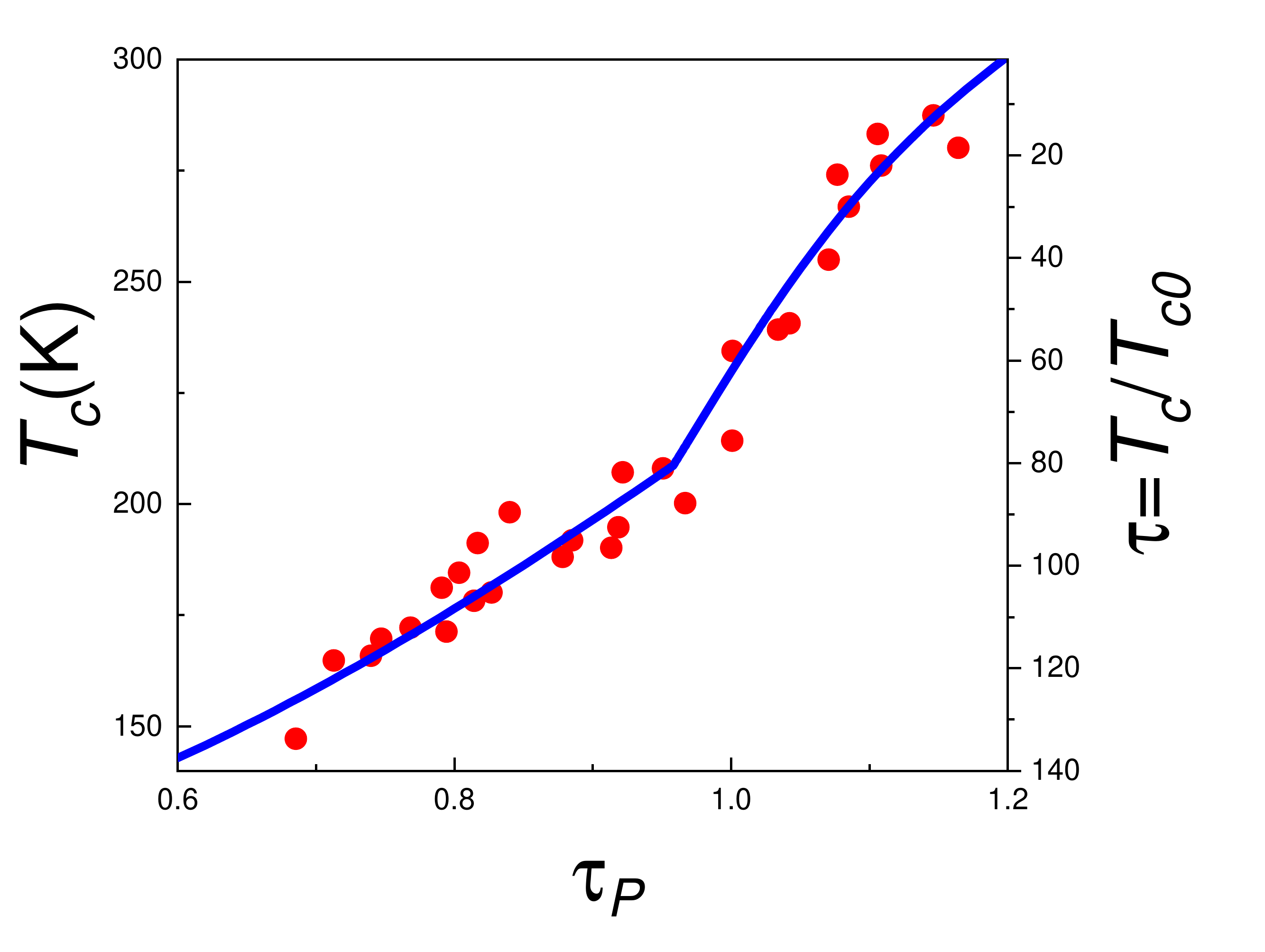}
    \caption{
Superconducting transition temperature as a function of $\tau_{P}$ for carbonaceous sulfur hydride CSH. Data points come from \citet{RN1}. log$\tau_{P}=K\Delta P$ where K is a constant and $\Delta P=P-P_0$ with $P_0$ chosen as the pressure where $T_{c0}$=233 K is observed. In the dirty limit $\tau_P\propto\tau$. The line in the figure is a fit using the GLAG theory in the cross over region as discussed in text. }
    \label{fig:6}
\end{figure}

\subsection{Understanding the Superconducting Transition Width}

From the GLAG theory, all thermodynamic measurements should exhibit noticeable variations within a temperature range where fluctuations of the order parameter exist. This will occur at the point in zero field where the extent of the wavefunction from the $G_{pair}$ term and wavefunction fluctuations from the $G_K$ term equal each other known as the Ginzburg criterion.\cite{RN11} This criterion yields the dimensionless Ginzburg or Ginzburg-Levanyuk (not Ginzburg-Landau) number $Gi_\text{clean}$ for an isotropic 3D superconductor in the clean limit given by:
\begin{equation}\label{eq:4}
    Gi_\text{clean}=\frac{1}{2}\left[\frac{ kT_{c}}{B^{2}_{c}\xi^{3}_{GL}}\right]^2
\end{equation}

Moving away from the clean limit and taking the result of Ferrell and Schmidt,\cite{RN12} the dirty limit value of the Ginzburg-Levanyuk number $Gi_{\text{dirty}}$ as:
\begin{equation}\label{eq:5}
Gi_{\text{dirty}}\approx\left[\frac{\xi_{GL}}{\ell}\right]^{2}\sqrt{Gi_{\text{clean}}}
\end{equation}

The value of $Gi$ physically gives the temperature range $Gi T_{c}$ near $T_{c}$ where fluctuations of the order parameter, and therefore to large changes in thermodynamic measurements, are observed. The transition width $\Delta T_{c}$ should be comparable to $GiT_{c}$. Rewriting gives $Gi=\Delta T_{c}$/$T_{c}$. While experimental $Gi$ values can be as low as $10^{-13}$ in clean Type-I 3D superconductors it is common for Type-II anisotropic superconductors to be in the $10^{-2}$ range as seen in MgB$_2$ and YBCO.\cite{RN11}

The superconducting transition widths decrease for increasing $T_{c}$ in the H-rich superconductors in contrast to what is expected, and observed, in superconductors in the clean limit. The $\tau$ dependence of $Gi$ in the clean limit is given by:
\begin{align}\label{eq:6}
\nonumber \frac{Gi_{\text{clean}}}{(Gi_{\text{clean}})_{0}}
&=\frac{\Delta T_{c}/T_{c}}{(\Delta T_{c}/T_{c})_{0}} \\
\nonumber {} &= \left[\frac{T_{c}/(T_{c})_{0}}{(B_c/(B_c)_0)^2(\xi_{GL}/(\xi_{GL})_0)^3}\right]^2 \\
\nonumber {} &= \left[\frac{\tau}{(\tau)^2(\tau^{-2})^3}\right]^2 \\
{} &= \tau^{10}
\end{align}

The $\tau$ dependence for the GLAG parameters here is taken from De Silva.\cite{RN8} This would lead to the conclusion that the superconducting transition increases rapidly with $T_{c}$ and effective mass as claimed by Hirsch and Marsiglio.\cite{RN2} Moving to the dirty limit, using the results from Figure 1, $Gi$ is:
\begin{align}\label{eq:7}
\nonumber\frac{Gi_{\text{dirty}}}{(Gi_{\text{dirty}})_{0}}
&=\left[\frac{\xi_{GL}/(\xi_{GL})_{0}}{\ell/\ell_{0}}\right]^2 \sqrt{\frac{Gi_{\text{clean}}}{(Gi_{\text{clean}})_{0}}} \\
\nonumber{}&=\left[\frac{\tau^{-2.1}}{\tau^{+1.9}}\right]^2\sqrt{\tau^{10}} \\
{}&=\tau^{-3} 
\end{align}

In the dirty limit this means that the superconducting transition will narrow with increasing $T_{c}$ as observed experimentally in heat capacity measurements on Nb$_{1-\beta}$Sn$_{\beta}$ samples.\cite{RN13}
For systems between the clean and dirty limits, the crossover region form $Gi_{cross}$ needs to be used where $\ell$ is replaced by $\xi_{AV}$. To further look at this scenario an examination of the transitions for all of the H-rich materials needs to be done. There is obviously going to be a great variation in the normal state zero temperature resistance as well as the zero field $\Delta T_{c}$ due to impurities, grain boundaries, disorder, pressure gradients, inhomogeneity and anisotropy to name just a few of many effects. To consistently examine literature values of $\Delta T_{c}$, the 75\%-50\% range of the maximum of all the selected normalized $R(T)$ data are used to exclude the broadening that often occurs at $T_{c}$(onset) and $T_{c}(R=0)$ due to multiple effects. This will obviously be a lower bound for the value of $\Delta T_{c}$, but this best exemplifies the majority phase (ideal) superconducting state. Also, many H-rich samples exhibit a 2 (or more) step transition at low fields that merge into a single transition that is related to the $T_{c}$ onset phase. Tinkham pointed out that the equations for the resistivity transition broadening do not depend on the range of temperature used as long as the range used is consistent.\cite{RN14} 

\begin{figure}[h]
    \includegraphics[width=\columnwidth]{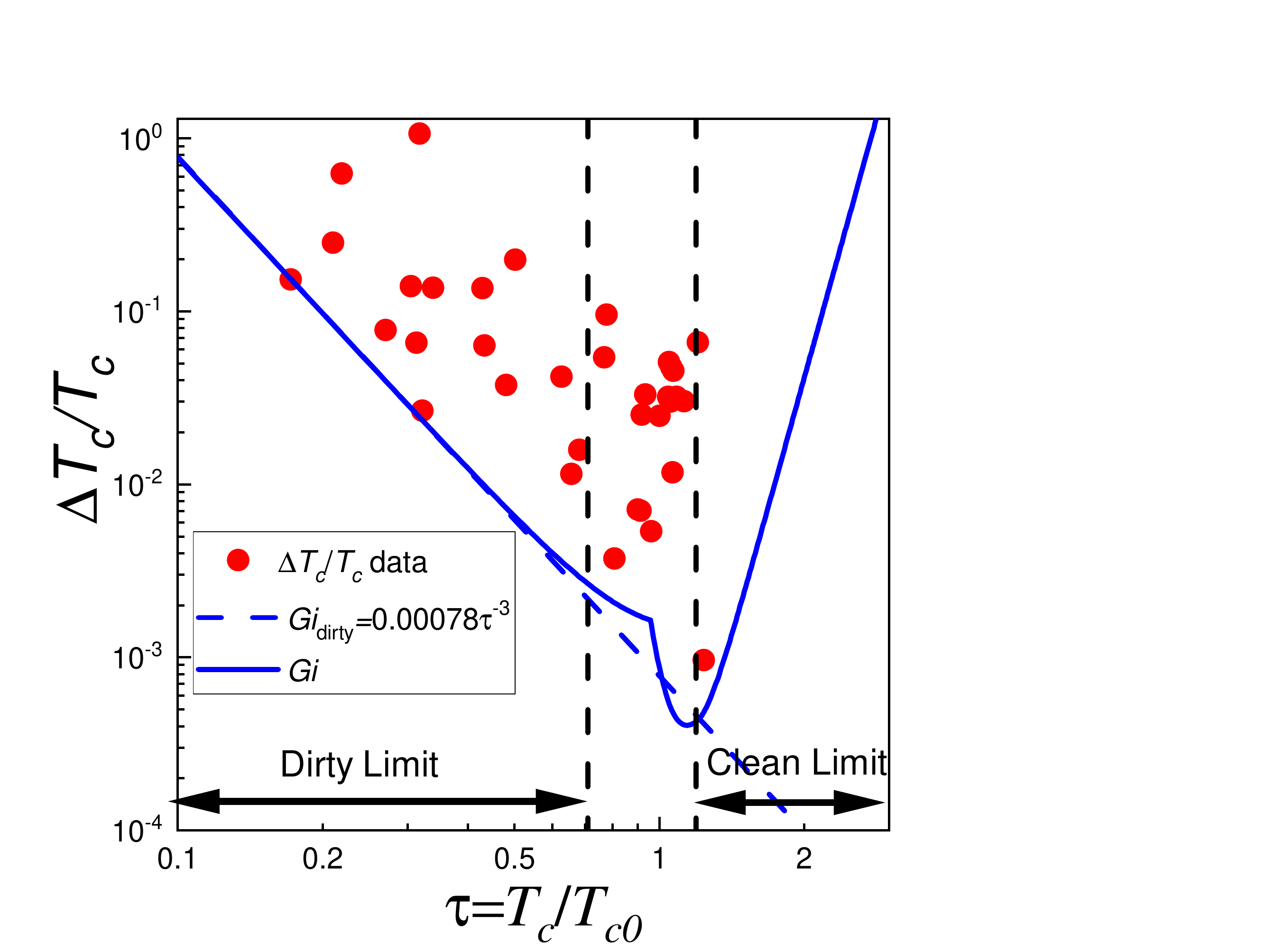}
    \caption{Values of $\Delta T_{c}/T_{c}$ for H-rich superconductors on a log-log plot. The dashed (solid) line are the Ginzburg-Levanyuk numbers $Gi$ in the dirty (cross over) regions that represent the minimum possible $\Delta T_{c}/T_{c}$ as a function of $\Delta T_{c}/T_{c0}$. The experimental values are all given in Table III. (inset) Experimental values of $\Delta T_{c}/T_{c}$ divided by the $Gi_{cross}$ line in Figure 7. This give a range of superconducting widths as a function of $\Delta T_{c}/T_{c0}$. Note that more or less dependent of $T_{c}$ the range of transition widths is consistent with a span around a factor of ~60 which is very close to the factor of 65 found for MgB$_2$ values in Table II.}
    \label{fig:7}
\end{figure}

The sharpest superconducting transition for a material will be given by the appropriate value of $Gi$. However, few systems will meet this ideal criteria. In reality, $\Delta T_{c}/T_{c}$ should go from $Gi$ to $Gi+\sum(\Delta T_{c}/T_{c}$) where the sum encompasses terms due to a number of factors (inhomogeneity/impurities, surface effects, anisotropy, thermal gradients, etc.). Table II shows numerous measurements on MgB$_2$ samples. While this is not a complete list, samples including bulk and films are included that display widely varying amounts of strain (including mechanically by ball milling) and impurities. These range of conditions should be a good starting point when comparing to the range of stress states in the H-rich materials. From the measured values in Table II, ($\Delta T_{c}/T_{c})_{min}$=0.0040. Compared to this minimum value, the largest broadening is just over 65 times larger. 

\begin{figure}[h]
    \includegraphics[width=\columnwidth]{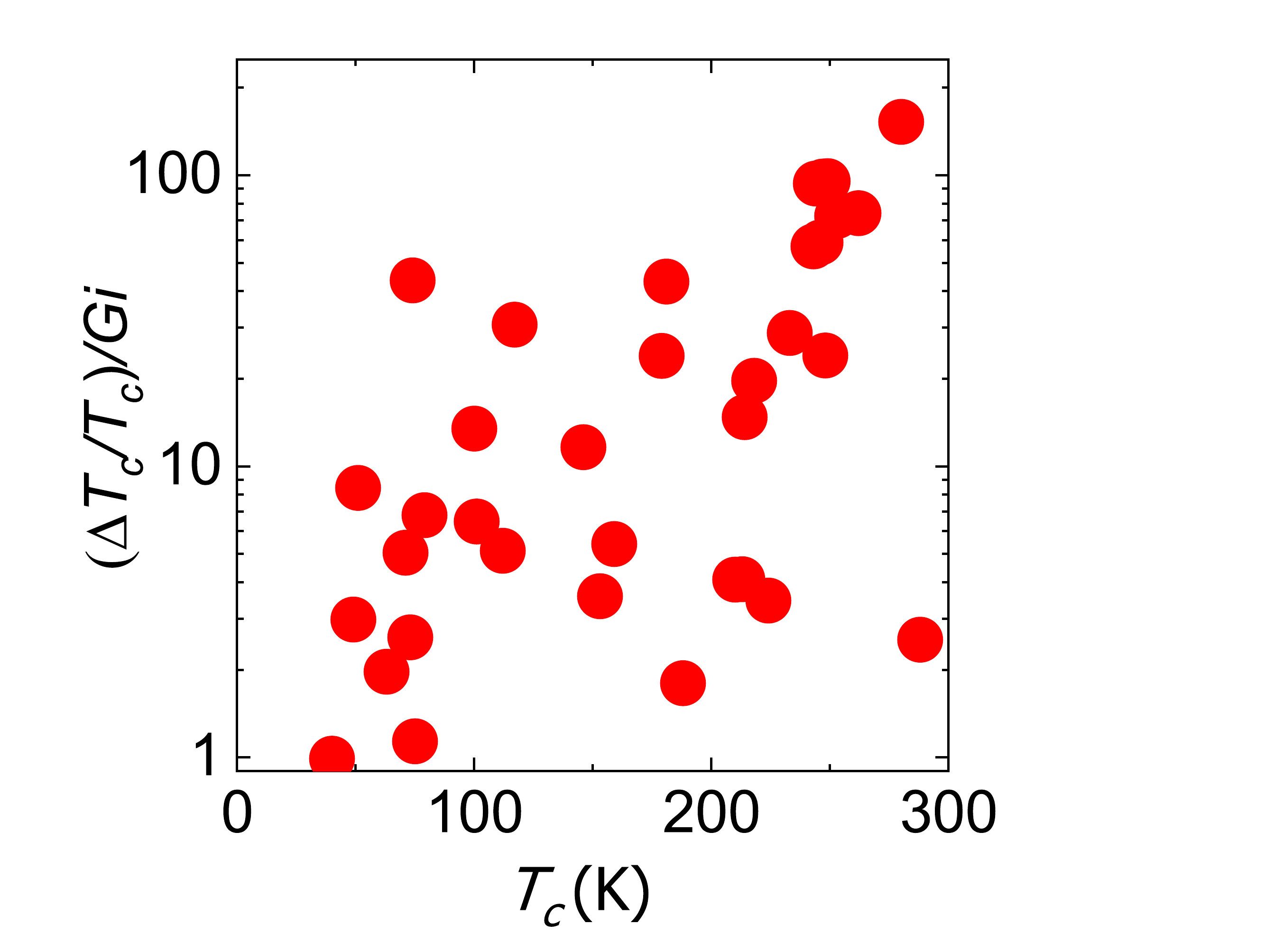}
    \caption{Experimental values of $\Delta T_c/T_{c0}$ divided by the $Gi$ line in Figure 7. This give a range of superconducting widths as a function of $\tau =T_c/T_{c0}$. Note that more or less independent of $T_c$ the range of transition widths is consistent with a span around a factor of ~60 which is very close to the factor of 65 found for Mg$B_2$ values in Table II. }
    \label{fig:8}
\end{figure}

Due to the great variation of the number of published high pressure data sets, criteria for choosing which data to analyze needs to be established. For those reports showing three or fewer $R(T)$ curves, all are analyzed. For those with more than three, the highest, lowest, and sharpest $T_{c}$ curves are analyzed. Also, any additional $R(T)$ measurements that include magnetic field data will be added and all results analyzed are given in Supplemental Table III. The broadening ($\Delta T_{c}/T_{c}$) versus $T_{c}$ in H-rich materials is shown in  Figure 7 on a log--log plot for all known H-rich superconductors. Looking at Figure 7, one can immediately see the trend that the transitions narrow with increasing $T_{c}$ values. The scatter at fixed values of $\tau$ is due to multiple already mentioned factors. The dashed line has the slope of $\tau^{-3}$ estimated from Equation 7 that goes through a single point for $\tau<0.65$ (this would correspond to the sharpest rescaled transition within the dirty limit). The solid line shows how $Gi$ should scale over the entire $\tau$ range using the $Gi_{cross}$ approximations from both limits. The two curves split very near 0.71 which is the end of the dirty limit and the start of the crossover region. Also $Gi_{cross}$ starts to increase dramatically as one move to the clean limit and gives the expected clean behavior that $Gi_C$ increases sharply with $T_{c}$.

\begin{figure}[h]
    \includegraphics[width=\columnwidth]{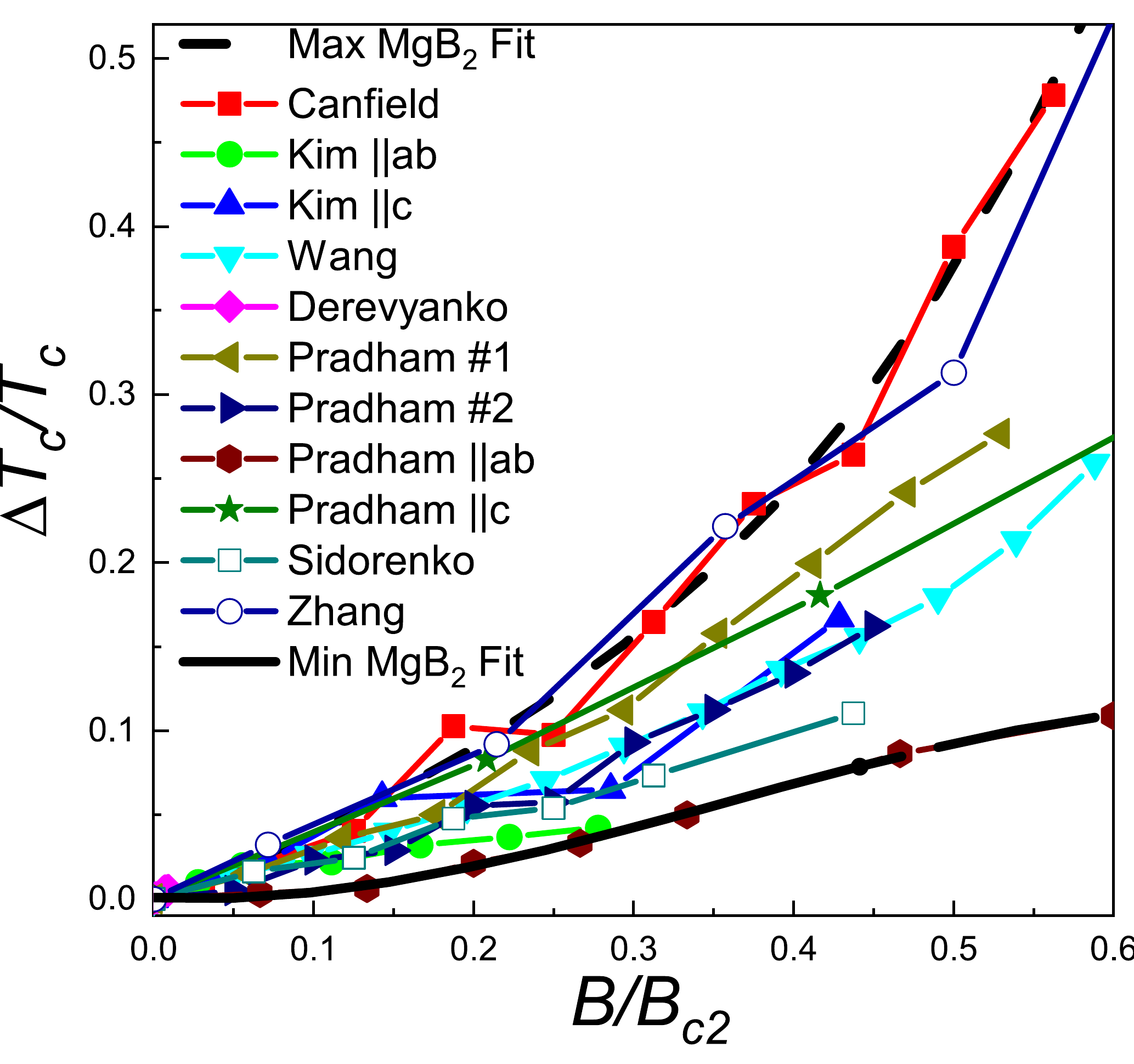}
    \caption{Compilation of numerous magnetic field broadening ($\Delta T_c/T_{c0})_B$ values for Mg$B_2$ as a function of normalized magnetic field $B/B_{c2}$. There are numerous single crystal, polycrystalline and film results. The references for each dataset are given in Table 1. }
    \label{fig:9}
\end{figure}

\begin{figure}[h!]
    \includegraphics[width=\columnwidth]{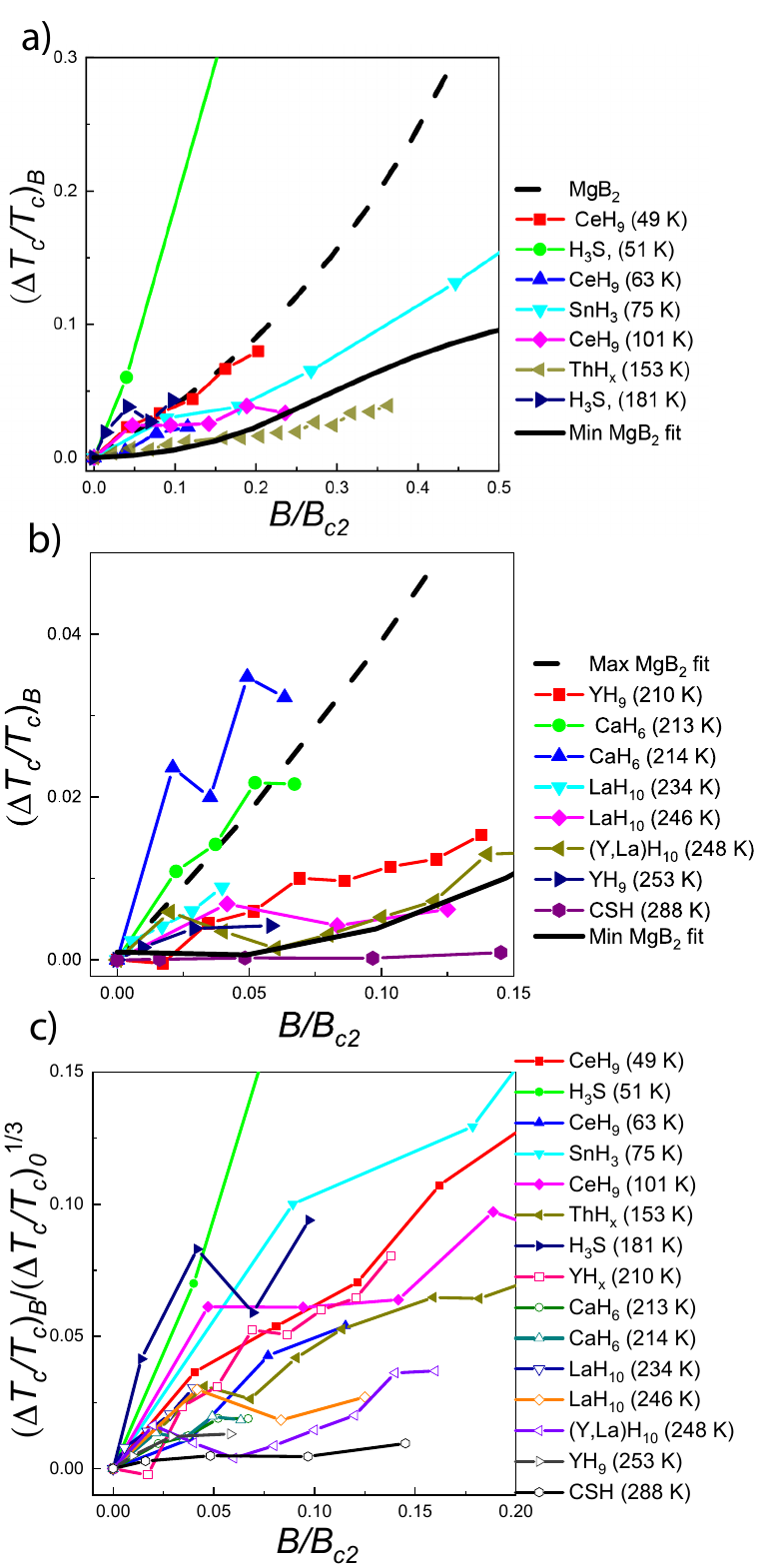}
    \caption {Values of $(\Delta T_c/T_c)_B$ for H-rich superconductors in applied magnetic field. The raw values for $T_c\,<$\,200\,K (a) and $T_c\,>$\,200\,K (b) are plotted along with the minimimum and maximum experimental values of MgB$_2$ from Figure 9. (c) The superconducting width rescaled by a factor of $((\Delta T_c/T_{c0})_0)^{1/3}$ which is the cube root of the zero field broadening.}
    \label{fig:10}
\end{figure}

Figure 8 shows the data in Figure 7 rescaled by the $Gi$ line from Figure 7. The range from minimum to maximum widths at all values of $\tau$ is very similar to the value of 65 for MgB$_2$. The fact that in the dirty limit the sharpest transitions scale as expected (get narrower as $\tau^{-3}$) and the spread in transition widths from smallest to largest is on the order of the range in values for MgB$_2$ naturally leads to the conclusion that the transition width behavior is far from anomalous and would be expected \textit{a priori} if starting all scaling from Figure 1.

\subsection{Broadening of Superconducting Transition in Applied Magnetic Field}

In applied magnetic fields superconducting transitions continue to broaden from their zero field value. Tinkham calculated that this magnetic field broadening should depend on $B^{2/3}$ in agreement with many experimental results.\cite{RN14} This result also follows from GLAG theory (Equation 1) as the addition of the $G_{pair}$ term in applied fields requires an extended temperature range for the mean field approximation to be valid. This leads to a field induced broadening of 
\begin{equation}\label{eq:8}
(\Delta T_c/T_c)_B=(Gi)^{1/3}(B/B_{c2})^{2/3},
\end{equation}
where $Gi$ is the zero-field Ginzburg number.\cite{RN38} As with zero field transition widths there are many factor that will increase ($\Delta T_c/T_c)_B$ from its minimum value in Equation 8 using $Gi$. This includes not only the conditions mentioned for zero field broadening but can include different modes of flux motion.

As with the zero field results, MgB$_2$ will be used as a starting point. The field dependent $(\Delta T_c/T_c)_B$  with the zero-field value of $(\Delta T_c/T_{c0})$ subtracted are shown for many MgB$_2$ measurements in Figure 9. The solid lines are 3rd order polynomial fits to the data of Canfield\cite{RN15} for a polycrystalline and  Pradhan\cite{RN16} for a single crystal sample. These two measurements are clearly a good representation of the range of the field dependent MgB$_2$ data. From Equation 8, the range of observed slopes should vary as the cube root of the zero field broadening, found to be $\sim$65 from Table I, which is approximately 4. For different values of $B/B_{c2}$ the ratio of the data of Canfield and Pradhan in Figure 9 vary by values from 3.5-5 in excellent agreement with this expectation.

Figure 10 shows the broadening for all of the H-rich supercondcutors for those with (a) $T_c$<200 K and (b)$T_c$>200 K separated here only for clarity. Most of the H-rich materials have values that lie within the range of the MgB$_2$ curves. There is a general trend that the slope decreases with higher $T_c$ value. This decrease in the magnetic field dependence is a natural result of Equation 8. However, there still exists a great deal of variation in the slopes of the H-rich data due to different broadening effects. To compare the H-rich samples for all $T_c$ values, the magnetic field data will be rescaled using Equation 8. However, instead of using the sharpest possible zero field broadening $Gi$, $Gi$ will be replaced by the zero field value of the broadening ($\Delta T_c/T_c)_0$ which should to some extent take into account the intrinsic broadening of the system.

Figure 10(c) shows the ($\Delta T_c/T_c$)$_B$ data from Figure 10(a) and 10(b) divided by a factor of  (($\Delta T_c/T_c$)$_0$)$^{1/3}$. The range of slopes now does not exhibit a general $T_c$ dependence and nothing stands out as anomalous.
To summarize the results on the transition broadening in H-rich materials:
\begin{itemize}
\item The range of $\Delta T_c/T_c$ values for the materials with similar $T_c$ values in zero field can vary greatly (around a factor of 60) for a variety of reasons. This range is independent of the value of $T_c$ and is similar to MgB$_2$.
\item 	The rescaled values of the additional broadening ($\Delta T_c/T_c$)$_B$ in applied fields show a wide variation as do the various values for MgB$_2$. The rescaled H-rich data show the expected transition broadening with applied magnetic field with neither a discernible $T_c$ dependence nor obvious anomalous behavior.\\
\end{itemize}

\subsection{Moving to the Clean Region}

The H-rich superconductors lie almost entirely outside of the clean limit. As great effort is being put into raising $T_{c}$, it is instructive to look at expected behavior from our GLAG scaling into this region. The highest $T_{c}$=288\,K in CSH puts all of the H-superconductors in the region $\tau$<1.24 which is just slightly into the clean region using our criteria. If $T_{c}$ is raised higher than 288\,K in H-rich systems the GLAG theory would predict the following. (1) The superconducting transition width will rapidly increase. (2) $B_{c2}$ should increase with increasing $T_{c}$. A minimum in $B_{c2}$ occurs due to the large increase in $\ell$ making $\xi=\xi_{GL}$ and $B_{c2}\propto(\xi_{GL})^{-2}\propto m^{*4}$. (3) $T_c$ will increase with an increase in effective mass. $T_{c}\propto m^{*7/3}$ which is faster than the linear behavior in the dirty limit.

\section{Conclusions}
Using the Ginzburg-Landau-Abrikosov-Gorkov (GLAG) theory, many aspects of the superconductivity in all of the hydrogen-rich materials observed to exhibit high-pressure superconductivity can be explained. This includes the following as $T_c$ increases: observation of a maximum in $B_{c2}$, a break in slope of $T_c(P)$, and decreasing superconducting transition widths. All of the H-rich superconductors follow the GLAG theory's predicted universal behavior regardless of applied pressure with effective mass increasing with $T_c$ being the driving force. This analysis also shows that raising the effective electron mass should continue raising the superconducting transition temperature even higher than room temperature. This should encourage more hydrogen dominant alloys to be synthesized that display larger effective masses and therefore $T_c$'s at lower densities.

\section{Acknowledgements}
This work supported in part by the U.S. Department of Energy, Office of Basic Energy Sciences under Award Number DE-SC0020303.

\section{Appendix A: Tables}
\begin{table*}[h!]
 \begin{tabular}{ |c||c|c|  } 
  \hline
&	H$_3$S &	LaH$_{10}$ \\
 \hline
$P$ (GPa) &	155	& 130\\ 
$T_c$ (K) &	196	 & 233 \\
$\tau=T_{c}/T_{c0}$	& 0.84 &	1.00 \\
WHH $B_{c2}$ (T) &	88 &	222 \\
$\xi$(nm) from $B_{c2}$	& 1.9 &	1.2 \\
$\ell$ (nm) from transport &	1.3 &	4.0 \\
$\xi_{AV}$ (nm) from $\xi$ and $\ell$	& 0.97 &	1.0 \\
$\xi_{GL}$ (nm) from $\xi$ and $\ell$	& 3.8	& 1.4 \\
$B_{c1}$ (T) &	1.7-2.5	& 0.8-3.0 \\
$\kappa$ from $B_{c1}$ and $B_{c2}$	& 5.5-7.1	& 9.0-20.5 \\
$\lambda_L=\kappa\xi_{AV}$ (nm)	& 5.3-6.9	& 9.0-20.5 \\
 \hline
\end{tabular}\\
    \caption{Ginzburg-Landau-Abrikosov-Gorkov (GLAG) parameters for H$_3$S and LaH$_{10}$ at the the given pressures. $T_{c} = 233 K$ is the value for the superconducting transition temperature in H-rich superconductors where $B_{c2}$ reaches its maximum value.} 
    \label{table1}
\end{table*}

\begin{table*}[h!]
 \begin{tabular}{ |c||c|c|c|c|  } 
  \hline
Sample Info	& $T_c$ (K)	& $\Delta T_c$ (K)	& $\Delta T_c/T_C$ &	Reference \\
  \hline
Poly &	40.2 &	0.49 &	0.012 &	Canfield\cite{RN15} \\
Single B|| &	36.0 &	0.32 &	0.0089 &	Kim\cite{RN19} \\
Single B||c	&36.0&	0.32&	0.0089&	Kim\cite{RN19}\\
Many doped MgB$_2$ Poly** &	37&	0.31&	0.0084&	Wang\cite{RN20} \\
Poly &	38.8 &	0.20&	0.0052&	Derevyanko\cite{Rn21}\\

Poly \#1 &	37.5	&0.36&	0.0096&	Pradhan\cite{RN16}\\

Poly \#2 &	38.0&	0.90&	0.023&	Pradhan\cite{RN16}\\

Single B||ab&	38.2&	0.69&	0.018&	Pradhan\cite{RN16}\\

Single B||c&	38.2&	0.69&	0.018&	Pradhan\cite{RN16}\\

TFilm&	36.2&	0.76&	0.021&	Sidorenko\cite{RN11}\\

TFilm&	40.2&	0.16&	0.0040&	Zhang\cite{RN22}\\

TFilm 600 C	&36.5&	9.5	&0.26&	Altin\cite{RN23}\\

TFilm 900 C	&36.5&	2.7	&0.074&	Altin\cite{RN23}\\

tFilm A	&36.2&	1.9	&0.054&	Altin\cite{RN24}\\

tFilm B	&33.3&	1.3	&0.038&	Altin\cite{RN24}\\

tFilm C	&33.4&	3.1	&0.091&	Altin\cite{RN24}\\

tFilm D&	31.2&	5.9	&0.19&	Altin\cite{RN24}\\

Poly 1h sinter&	38.5&	0.59&	0.015&	Mizutani\cite{RN25}\\

Poly 6h sinter	&38.4&	0.57&	0.014&	Mizutani\cite{RN25}\\

Poly 24h sinter	&38.4&	0.48&	0.013&	Mizutani\cite{RN25}\\

tFilm&	39.2&	0.21&	0.0054&	Kang\cite{RN26}\\
\hline
\end{tabular}\\
    \caption{Superconducting properties of MgB$_2$ samples from the literature having magnetic field measurements. For sample info: Poly and Single mean polycrystalline and single crystal respectively. tFilm and TFilm mean thin and thick film respectively. Any other text in the Sample Info column is descriptive information from the authors. **Values are for 8 pure and doped polycrystalline samples. } 
    \label{table2}
\end{table*}

\begin{table*}[h!]
 \begin{tabular}{|c||c|c|c|c|c|c|} 
  \hline
System	& $T_c$ (K) &	$P$ (GPa) &	$\Delta T_c$ (K) &	$\Delta T_c /T_c$ &	$B_{c2}$(T) (GL/WHH) &	Reference\\
  \hline
CeH$_9$ & 49	&88&	12.2&	0.24&	24.7/33.5	&Chen\cite{RN27}\\

&	101	&139	&6.4&	0.063&	21.2/28.6&\\	
&	63	&100	&4.9&	0.078&	26.0/35.6&\\
  \hline
YH$_x$&	224&	166	&1.2	&0.0055&	N/A	&Troyan\cite{RN28}\\

&	218	&165	&7.2&	0.033&	N/A	&\\
&	210	&183	&1.5&	0.0070&	116/158&\\	
  \hline
H$_3$S&	51&	155&	31.9&	0.64&	25/33*&	Drozdov\cite{RN6}\\

&	181&	195	&17.3&	0.096&	72/85*	&\\
&	21	&107	&9.7&	0.46&	&	\\
&	71	&177	&9.9&	0.14&	&	\\
  \hline
LaH$_{10}$	&246	&150	&7.4	&0.030&	115/160*&	Drozdov\cite{RN29}\\

&	112	&150	&4.2&	0.038&&\\		
&	73	&150&	4.8	&0.066	&&\\
  \hline
CSH	&280&	272&	18.5&	0.068&&		Snider\cite{RN1}\\
&	188	&210	&0.7&	0.0032&	48/66&	\\
&	288	&267	&0.28&	0.00010&	62/85&	\\
  \hline
LaH$_{10}$&	248	&188&	2.9&	0.012&	N/A&	Somayazulu\cite{RN30}\\
  \hline
LaH$_{10}$	&233&	165&	5.8&	0.025&	176/222&	Hong\cite{RN31}\\

&	74	&165&	78.8&	1.06&	N/A	&\\
  \hline
SnH$_x$&	75&	200&	2.0	&0.026	&11.2/9.4&	Hong\cite{RN32}\\
  \hline
(La,Y)H$_{10}$	&249&	186&	11.3	&0.046&	100/135&	Semenok\cite{RN33}\\

&	243	&191&	7.76&	0.032&	N/A&\\	
&	247	&183&	11.7&	0.048&	N/A	&\\
  \hline
ThH$_x$&	153&	170	&1.75&	0.012&	44/63&	Semenok\cite{RN34}\\

&	159	&174&	2.51&	0.016&	N/A&\\	
&	146	&170&	6.1	&0.038&	N/A	&\\
  \hline
PH$_3$	&100&	207&	13.6&	0.137&	N/A&	Drozdov\cite{RN35}\\

&	79	&180	&10.8&	0.138&	N/A&\\	
&	40	&117	&6.1&	0.152&	N/A	&\\
  \hline
YH$_x$&	262	&182&	7.9&	0.030&	N/A	&Snider\cite{RN36}\\

&	253	&177&	8.1&	0.032&	102/117*	&\\
&	244	&144&	12.4&	0.047&	N/A	&\\
  \hline
CaH$_6$&	117&	139	&23.2&	0.198&	N/A	&Ma\cite{RN37}\\

&	179	&129&	9.7&	0.054&	N/A	&\\
&	213	&172&	1.5&	0.0070&	131/180	&\\              
&	214	&178&	5.4&	0.025&	142/203	&\\
 \hline
\end{tabular}\\
     \caption{Superconducting parameters for H-rich materials at high pressure. The $B_{c2}$ values were taken or calculated from the referenced publication. GL/WHH refer to the Ginzburg-Landau (GL) and Werthamer-Helfand-Hohenberg (WHH) formalisms for estimating the zero-temperature value $B_{c2}$= $B_{c2}$(0). * Value was estimated from the published data for the WHH formalism using $B_{c2}=-0.69 T_c B_{c2}\prime$ where $B_{c2}\prime$ is the temperature derivative of $B_{c2}$ at $T_c$. } 
    \label{table3}
\end{table*}


%
\end{document}